\documentclass[sigconf]{acmart}
\usepackage{amsmath,amssymb,amsfonts}
\usepackage{moreverb,algorithm,algorithmic}
\usepackage{graphicx}
\usepackage{caption}

\setcopyright{acmcopyright}
\copyrightyear{2019}
\acmYear{2019}
\acmDOI{10.1145/3366486.3366488}

\acmConference[To Appear in WUWNET'19]{WUWNET'19: International Conference on Underwater Networks and Systems}{October 23--25, 2019}{Atlanta, GA, USA}

\begin{document}

\title{Compressed Underwater Acoustic Communications for Dynamic Interaction with Underwater Vehicles}

\author{
Mehdi Rahmati, Archana Arjula, and Dario Pompili}
\affiliation{Department of Electrical and Computer Engineering, Rutgers University--New Brunswick, NJ, USA}
\email{{mehdi.rahmati, archana.arjula, pompili}@rutgers.edu}

\begin{abstract}\label{sec:abstract}
Underwater vehicles are utilized in various applications including underwater data-collection missions. The tethered connection constrains the mission both in distance traveled and number of vehicles that can run in the same area, while the addition of acoustic communications onto the vehicles grants them several functionalities. 
However, due to the low bandwidth of the underwater acoustic channel---which leads to low data rates---and the time overhead imposed by both the channel propagation delay and the processing delay by the acoustic modems, efficient protocols are required. In this paper, 
an implicit data-compression and transmission protocol is proposed to carry out environmental monitoring missions such as adaptive sampling of physical and chemical parameters in the water. In a semi-autonomous manner between the vehicle and the control center, both sides keep silent in data transmission as long as they can estimate and predict the actions of the other side, unless environmental data and/or kinematic data are found to be unpredictable. Our design puts the human in the loop to send high-level control commands. Experiments were conducted using an autonomous vehicle with WHOI micro-modems in the Raritan River, Somerset, Carnegie Lake in Princeton, and in the Marine Park in Red Bank, all in New Jersey.
\end{abstract}

\maketitle

\section{Introduction}\label{sec:intro}

\textbf{Overview:}
Underwater communications and networks---using static nodes and/or mobile vehicles---enable a wide range of applications such as oceanographic data collection, pollution monitoring, marine life imaging, disaster prevention, and tactical surveillance~\cite{rahmati2017unisec}. In many of these applications, Autonomous Underwater Vehicles~(AUVs) or Remotely Operated Vehicles~(ROVs), equipped with multiple on-board sensors, are used in the exploration and exploitation of undersea resources for gathering scientific data in collaborative monitoring missions~\cite{rahmati2019network}. 
Currently underwater vehicles are often tethered to the supporting ship by a cable or have to surface periodically to communicate with a remote onshore station via terrestrial Radio Frequency~(RF) communications. Tethering is a serious limitation for the development of underwater systems. 

\textbf{Challenges:} 
Although acoustic communication is the typical technology underwater for distances above a hundred meters, yet, achieving high data rates transmission through the acoustic channel is hard to accomplish as acoustic waves suffer from attenuation, limited bandwidth, Doppler spreading, high propagation delay, and time-varying channel characteristics~\cite{rahmati2017ssfb}. 
Moreover, timely reaction to the control commands is a serious challenge in the time-critical applications due to the huge amount of delay caused by the propagation delay and the delay in the operation of acoustic modems. Regarding different levels of autonomy for controlling an underwater vehicle, full-autonomy status is not always feasible due to practical issues in the water including the navigation/localization errors. Therefore, using a communication system in a user-assisted autonomy status, i.e., semi-autonomy, is another challenge that we would like to tackle in this paper. 


\textbf{Related Work:}
Adaptive sampling is a strategy for data collection from the places where the data is more valuable to collect~\cite{rahmati2018slam}. Authors in~\cite{chen2010trajectory} employ a trajectory-aware solution by using geocasting within an estimated position uncertainty region. 
The uncertainty region is modeled for networked autonomous vehicles as in~\cite{rahmati2015interference}. An Unscented Kalman Filter~(UKF) is applied to the uncertainty model to improve the precision and to reduce the uncertainty of the model~\cite{rahmati2018probabilistic}. 
Authors in~\cite{5066973} use Kalman filtering to approximate the trajectory of a vehicle, such that the vehicle only have to report its trajectory when there is an adjustment in it. 
Authors in~\cite{zhu2005challenges} propose an energy efficient communication strategy for wireless sensor networks to convey information between the nodes using silence periods between the energy signals. 

The WHOI micro-modem~\cite{freitag2005whoi} has been implemented in vehicles such as Nereus in~\cite{singh2009acoustic}, which can operate as an AUV or a tethered ROV. This is similar to our design; however, it opts for the tethered link for its ROV mode and was designed to work in deep-ocean settings with vertical communication links. The versatility of the WHOI micro-modems is important which is able to operate in both deep vertical links and shallow water horizontal links environments. The Nereus experiments also display the long-range capabilities of the micro-modem by being able to communicate across a distance of $11~\rm{Km}$, which reveals that these micro-modems are strong enough for different applications. Authors in~\cite{marques2007auv} describe the use of acoustic networks for data collection in AUVs via a control and communication framework. The framework uses both RF and acoustic modems to enable long range and high rate access and aims at reducing the transmission time and errors. In~\cite{curtin2009progress}, a decision-making process is introduced for the received data to make prediction of the dynamic ocean fields. 


\textbf{Our Contribution:}
We propose a hardware design and a communication protocol for an underwater vehicle---which is capable of performing some local autonomous controls---to send/receive data/command to/from an outside control center in order to collect environmental underwater data. Using this protocol, the vehicle can be pre-configured to perform various different types of missions in order to achieve the ability to decide quickly to analyze the measured valuable data (or unpredictable change in the pre-configured settings) locally or report it to the control center for further commands. 
Our design introduces a high-level control which puts the human in the loop for further decisions in the semi-autonomous mode. 
To accomplish this goal, after full installation of a WHOI modem on an underwater ROV, we propose a compression and communications protocol for the data---which is not predictable by the other side of the link---considering the constraints of underwater acoustic modems. 
Due to the low bandwidth of the underwater acoustic channel---which leads to a low data rate---and the time overhead imposed by both the channel propagation delay and the acoustic modems processing delay, vehicle will not transmit the data as long as there is a matching between the vehicle and the control center. 
Since the inclusion of the semi-autonomous mode faces the low throughput and high delay of the acoustic channel and the micro-modems, we aim to resolve this problem through efficient solutions of compression and selective transmission methodology. 
\section{System Models and Our Solution}\label{sec:solution}
In this section, first, 
we discuss different types of data that we can collect during the tests from the field. 
Afterwards, we explain the operation modes and necessity of data compression and our solution in order to transmit the data effectively. 
Finally, 
we describe the data communications protocol that we propose to transmit the unpredictable data using the WHOI micro-modems.

\textbf{Data Collection and Adaptive Sampling:} 
We consider a scenario in which a single vehicle performs sampling for near-real-time water-quality monitoring in a river, lake, or a water reservoir; the solution can be extended to a multi-vehicle scenario. To sample a region of interest, the vehicle moves around the region by following a certain trajectory and takes samples. The measured temperature, as a sample data, helps the controller of the vehicle decide whether to move towards or away from a warmer/colder area depending on the application and mission goal. The dissolved oxygen level---which is the amount of free, non-compound oxygen in water bodies---is also an important parameter obtained to help observe the habitat of marine animals/plants.
%
%
%
%
%
The location data that indicates the Global Positioning System~(GPS) coordinates---when the vehicle is on the surface---is also important to send back to the control station so that the user on the other side knows where the vehicle is located. The other kinematic properties of the robot, such as the sudden change in acceleration and orientation, i.e., magnetometer and gyroscope sensor, could also be collected through the sensor readings from the Inertial Measurement Unit~(IMU). The pressure data, collected from the barometer, will also be used to estimate the depth.  

\textbf{Operation Modes:} 
In the \textit{autonomous mode}, the vehicle is given a mission which is programmed on-shore. The vehicle will know how to execute this mission and make any adjustments through local processing in order to complete the mission as it is programmed. No wireless underwater acoustic communication is used in this method. The second mode, which is the \textit{semi-autonomous mode}, 
involves the use of underwater acoustic wireless communication to either reprogram the mission, change some of the parameters while the vehicle is underwater, or make the vehicle stop its mission early and come back to shore. 
When compared to fully autonomous modes, the semi-autonomous vehicles are considered superior due to the fact that they can respond to the intermediate commands from the control center. In the semi-autonomous mode, the vehicle could be reprogrammed from a mission to perform a different task in the middle of a mission, performing an entirely new mission after the completion of another one without returning to shore, and changing the parameters of the vehicle's control system during a mission. These are usually done when there is the possibility of data transmission from vehicle to the other side of the link. 

\begin{algorithm}[!t]
\caption{Selecting optimal data points} \label{algo:compress}
 \small
 \begin{flushleft}
  \textbf{Input (X):} {Sensor measurements;\;} 
 \textbf{Output (Y):} {Compressed data points} \\
 \end{flushleft}
 \begin{algorithmic}[1] \small
 \STATE{\textit{Model-free approach for environmental data}} \\
  \REPEAT 
    \STATE{Compute $\mu$ from the data given}
    \STATE{Estimate the least square error of descriptors of X from the cutoff $\mu$}
   \UNTIL{|$X$ $-$ $\mu$|$^2$ $>$ $T_h$ \hspace{20mm} \% $T_h$: Error Threshold }
 \STATE{\textit{Model-based approach for kinematic data}}
 \WHILE{Algorithm is running}
 \STATE{\textbf{Prediction:} Predict $x_{t+1}$ from previous step $x_{t}$ using~\eqref{eq:1}}
 \STATE{Calculate the covariance from~\eqref{eq:1}}
 \STATE{\textbf{Updating:} Get the Kalman Filter Gain $K_{t+1}$ from~\eqref{eq:3}}
 \STATE{Update the next step $x_{t}^{'}$ given $x_{t+1}$ using Kalman gain}
 \STATE{Get $x_{t}^{'}$ from Kalman filter; compare with sensor measurements X}
 \STATE{\textbf{if} result differs from the measurements by a threshold\\ \textbf{then}} Send those points using Algorithm (2)
 \ENDWHILE
\end{algorithmic}
\end{algorithm}

\textbf{Data Compression:}
Sending all the data collected by the vehicle back to the shore might cause a high delay given the limited bandwidth and high propagation delay of the acoustic channel. Therefore, data needs to be compressed before transmission by eliminating the redundant data and by choosing the relevant data to transmit without causing any disturbance to the mission. 
The solution includes creating a decision scheme that limits the kind of information is being sent through the channel. There are two general categories of data compression as explained bellow.

\begin{figure*}[!t]
\centering
\vspace{-3mm}
\includegraphics [width=0.96\textwidth]{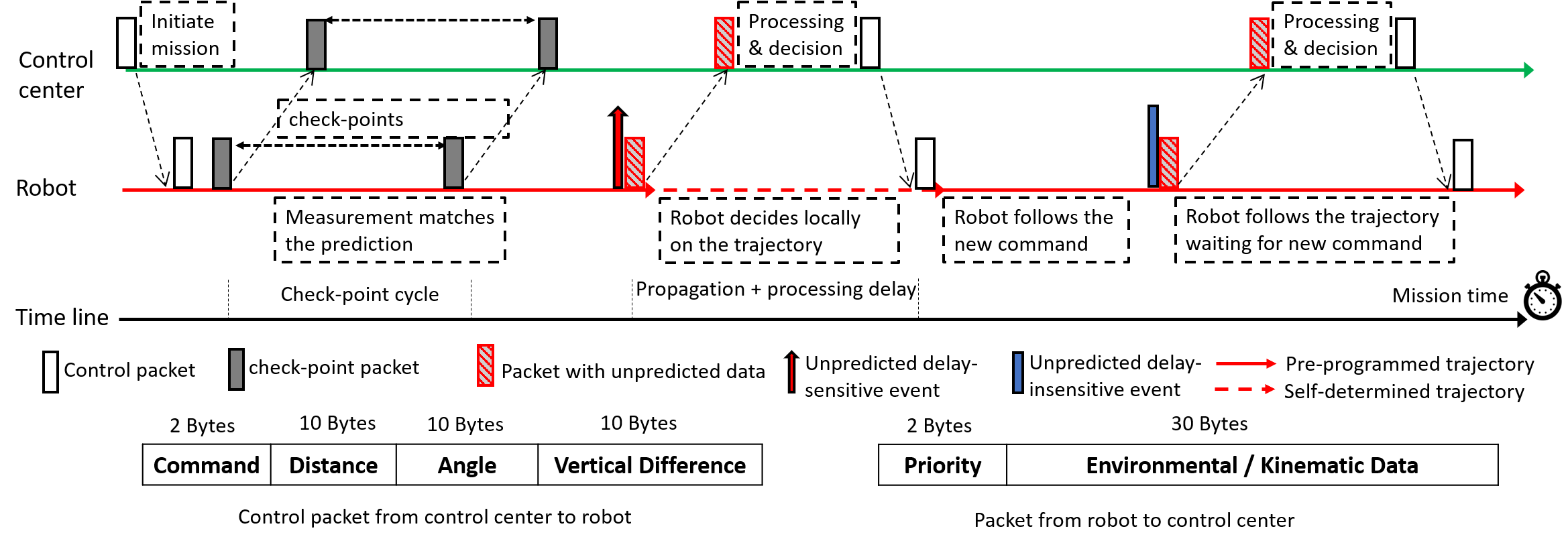}
\vspace{-2mm}
\caption{Timeline of the interaction between robot and the control center. While the measured data matches the predicted one, the robot remains silent and only sends check-point packets. In the case that any unpredicted event happens, the robot sends a priority packet containing the new environmental/kinematic data. Predicted or self-determined trajectories are followed based on sensitivity or insensitivity of the event.}\label{fig:timeline}
\vspace{-3mm}
\end{figure*}

\textit{Environmental data compression:} 
Based on the application and regarding the information-theory definition of the valuable data, this type of data considers any abrupt increase/decrease in the expected data points, i.e., a high entropy as a measure of uncertainty. Further actions might be required if the vehicle needs to change his direction towards those regions with more valuable data. When the change in temperature is noted, the starting and ending data points are selected (intermediate points could be selected as well). Algorithms, such as Gaussian Process~(GP) or Fast Marching~\cite{sethian1996fast}, can be applied to find the boundary of the valuable data around the point which was initially found. 


\textit{Kinematic data compression:}
In the process of compression of IMU data, the most unpredictable data should be prioritized. Kalman filtering is used for vehicle tracking and trajectory prediction from the collected IMU data. If the predicted data points are close to the measured data then those data points are ignored. Otherwise, the data is labeled as unpredicted data. In the first step, robot's position $x$ at time step $t+1$ is predicted based on its current location at time $t$ and its movement due to the control input $u_{t}$ as follows,
\begin{equation} \label{eq:1}
x_{t+1} = F_{t}x_{t}+ B_{t}u_{t}, \;\;\;
P_{t+1} = F_{t}P_{t}F_{t}^T+ Q_{t},
\end{equation}
%
where $F$ is the state transition matrix, $P$ is the a-priori covariance matrix, $u$ is the control variable, $B$ is the control matrix and $Q$ is the process variance matrix (i.e error in the process). In the second step of observation, the measurements $z_{t+1}$ are recorded from the vehicle's sensors at the location during time $t+1$. The observations consists of a set of single state elements extracted from different sensors. In the next step, it predicts the position of the robot and the map to generate the predicted observations, which are transformed from the sensor frame,
\begin{equation} \label{eq:2}
z_{t+1} = H_{t+1}+ x_{t+1},
\end{equation}
where $H$ is the state transition model. By combining both the prediction of robot position $x_{t+1}$ with the Kalman gain observed by the measurements, we get the updated estimate of the robot's position as follows,
\begin{equation}\label{eq:3}
    K_{t+1} = P_{t+1}H_{t+1}^{T}[H_{t}P_{t}H_{t}^T + R], \:\,
    x_{t}^{'} = x_{t+1}+ K_{t+1}z_{t+1},
\end{equation}
%
%
where $R$ is the measurement of noise, $x_{t}^{'}$ is the final updated estimate calculated from the $K_{t+1}^{'}$ which is the Kalman gain. 
If the vehicle's IMU data does not coincide with the estimations from Kalman filter, then the data points are marked and sent to the control center through WHOI modems, i.e., acoustic communications. Algorithm~\ref{algo:compress} explains these steps in more details.

\textbf{Data Communications Protocol:}
We create a script and a protocol to initialize the communication link, receive a command from the control center, and transmit data using a specified packet format over the communication link for the underwater vehicle. We define two formats for our packets, i.e., control and data packets, where the former are loss-sensitive and could be either delay-sensitive or -insensitive depending on the criticality of the command. 
%

Fig.~\ref{fig:timeline} shows the packet format in each direction. Control packet includes command, distance, angle, and vertical difference which refer to how far the vehicle moves, the change in angle when the vehicle turns, and difference in the $z$ direction, i.e., depth, respectively. The packet sent from robot specifies two bytes to identify the priority of the data and 30 bytes for environmental/kinematic data. Each packet has $32$ bytes to match type $0$ packet of WHOI as will be explained in Sect.~\ref{sec:Elav}. Fig.~\ref{fig:timeline} also represents the timeline for the proposed communication protocol in the semi-autonomous mode. The control center initiates the mission by sending the proper control command and in response the robot transmits the check-point packet. 
While the measured data matches the predicted one, the robot remains silent and just sends check-point packets to keep synced with the center. In the case that any unpredicted event happens, the robot sends a priority packet containing the new environmental/kinematic data. If the data is delay-sensitive and needs an urgent consideration, the robot decides locally to whether deviate its trajectory towards the source of the event. Meanwhile the control center receives the updated information and makes decision on the mission whether the robot follows a new trajectory or returns to its original path. Under the reception of the new command in control packet, the robot reacts accordingly. In the case that the data is delay-insensitive, the vehicle follows its original trajectory while waiting for the control center's decision. Algorithm~\ref{algo:comprotocol} explains the procedure in details.             




\begin{algorithm}[!t]
\caption{Comms protocol in semi-autonomous mode} \label{algo:comprotocol}
 \small
 \begin{algorithmic}[1] \small
    \STATE{Initialization; Wait until control packet is received from the center }
  \STATE{Send the check-point to the center; mission started}
   \WHILE{data $x\approx x'$ OR $t < T_{\rm{mission}}$}
 \STATE{Continue the pre-programmed trajectory; Send the checkpoint}
 \ENDWHILE
 \IF{$t > T_{\rm{mission}}$}
 \STATE{Go to End (back to the center)}
 \ELSIF{the data is time-sensitive}
 \STATE{Decide locally on trajectory; Send priority packet to control center} 
 \REPEAT 
    \STATE{Follow the self-determined trajectory}
   \UNTIL{receive control packet from the center}
   \STATE{Follow the control command}
   \ELSIF{the data is time-insensitive}
   \STATE{Send priority packet to control center}
   \REPEAT
   \STATE{Continue the pre-programmed trajectory}
   \UNTIL{receive control packet from the center}
   \ENDIF
   \IF{$t < T_{\rm{mission}}$}
   \STATE{Go to line $3$}
   \ENDIF 
 \end{algorithmic}
\end{algorithm}

\section{Performance Evaluation}\label{sec:Elav}
First, we focus on the hardware installation of the WHOI micro-modems into our vehicle. The protocol is defined using National Marine Electronics Association~(NMEA) commands over the WHOI micro-modems and python script on the vehicle's processor. Then, we implement the communication protocol and benchmark the performance through controlled tests using our proposed solution. 

\textbf{Vehicle Development with Micro Modems:}
The WHOI acoustic micro-modem has different packet types, modulation schemes, number of bytes in the frames, maximum frames per packet, and their packet payload in terms of bps at $5~\rm{KHz}$ bandwidth~\cite{whoisoftware}. The effectiveness of a packet type relies on the region and on the underwater acoustic channel. Packet Type $0$ is the least error prone type in low SINRs but the throughput of Packet Type $5$ is the highest.  
The decision on which packet type to use is taken based on the underwater channel along with a metric between loss-sensitivity and delay-sensitivity of the application. 
\begin{figure}[!t]
\centering
\vspace{-2mm}
\includegraphics [width=0.48\textwidth]{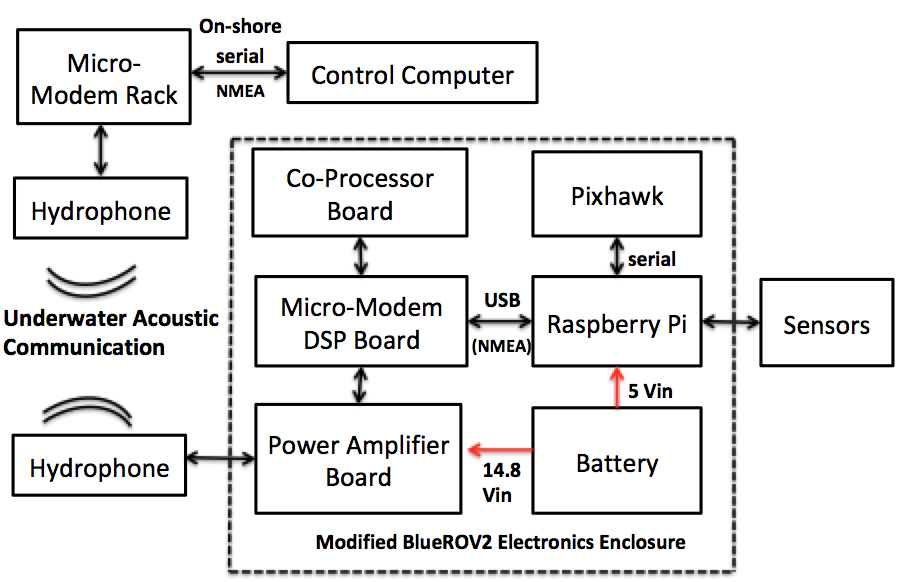}
\caption{Block diagram of the modified vehicle for the shore (control center) to vehicle acoustic communication.}\label{fig:block}
\vspace{-3mm}
\end{figure}
%
Type $0$ packets have the smallest data payload size of $32$ Bytes. As this type is usually the least error prone (depending on the SINR), it is our most widely used packet as we are building a communication system under the assumption that it can operate in the worst-case scenario. Thus the $32$ Byte payload will be the upper bound for our data payload length in a single packet. While using packet types with larger frame sizes, our format will still be limited to a $32$ Bytes payload in order to establish uniformity between the packets when being transmitted over different rates and schemes.

Going over the IMU data collected in the experiments, assume there are seven data points that are $16$ bits each being collected every $0.25$ seconds. Therefore, $56$ Bytes should be transmitted every second to not incur a loss. As the micro-modem has to transmit in cycles in order to process signals coming in on all of its $4$ channels due to multipath, we need a packet type that has a large datarate compared to frame-size so that it will be above the lower limit of $56$ Bytes per second threshold. We can alter the sampling rate of the IMU to better fit with a suitable packet size for data transmission which does not have such a large throughput requirement if needed. 
Using NMEA commands, a data payload is given to the transmitter along with a source, destination, and whether or not an ACK is required. As expected, the slowest packet type is type $0$. The data being transmitted with this format is as small as a $32$ byte payload. Packet type $3$ has a large frame size which is about four times as large as packet type $2$; however, the data rate difference between types $2$ and $3$ is only about $2.4$ times. Thus, packet type $3$ has to send a larger frame which is mostly blank at a rate which is better but it does not offset the frame size compared to packet type $2$. 


\begin{figure}[t!]
\centering
\begin{tabular}{cc}
\hspace{-4mm}
\includegraphics [width=0.16\textwidth]{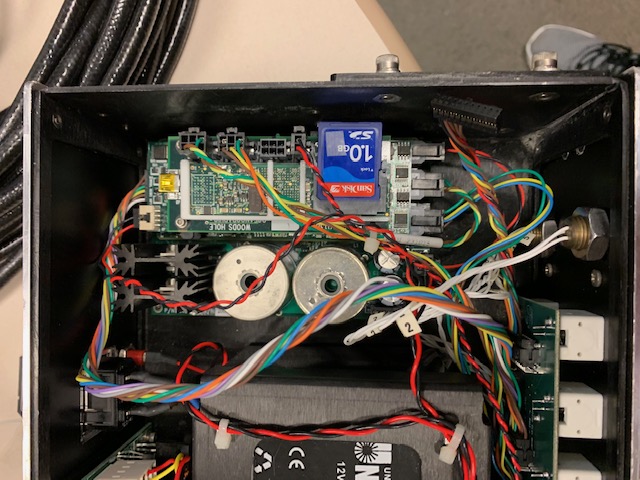}
\includegraphics [width=0.16\textwidth]{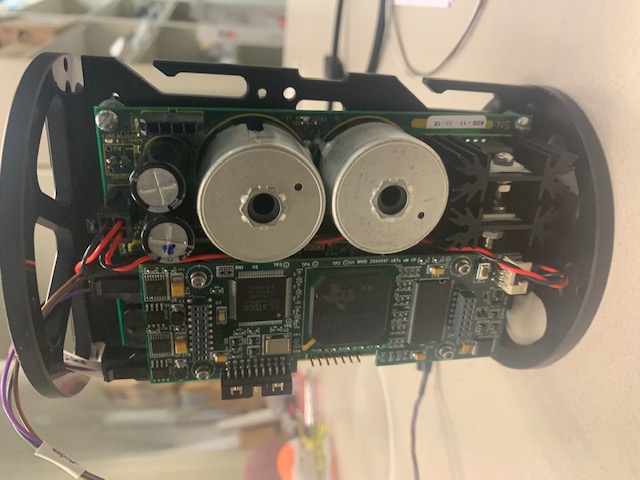}
\includegraphics [width=0.17\textwidth]{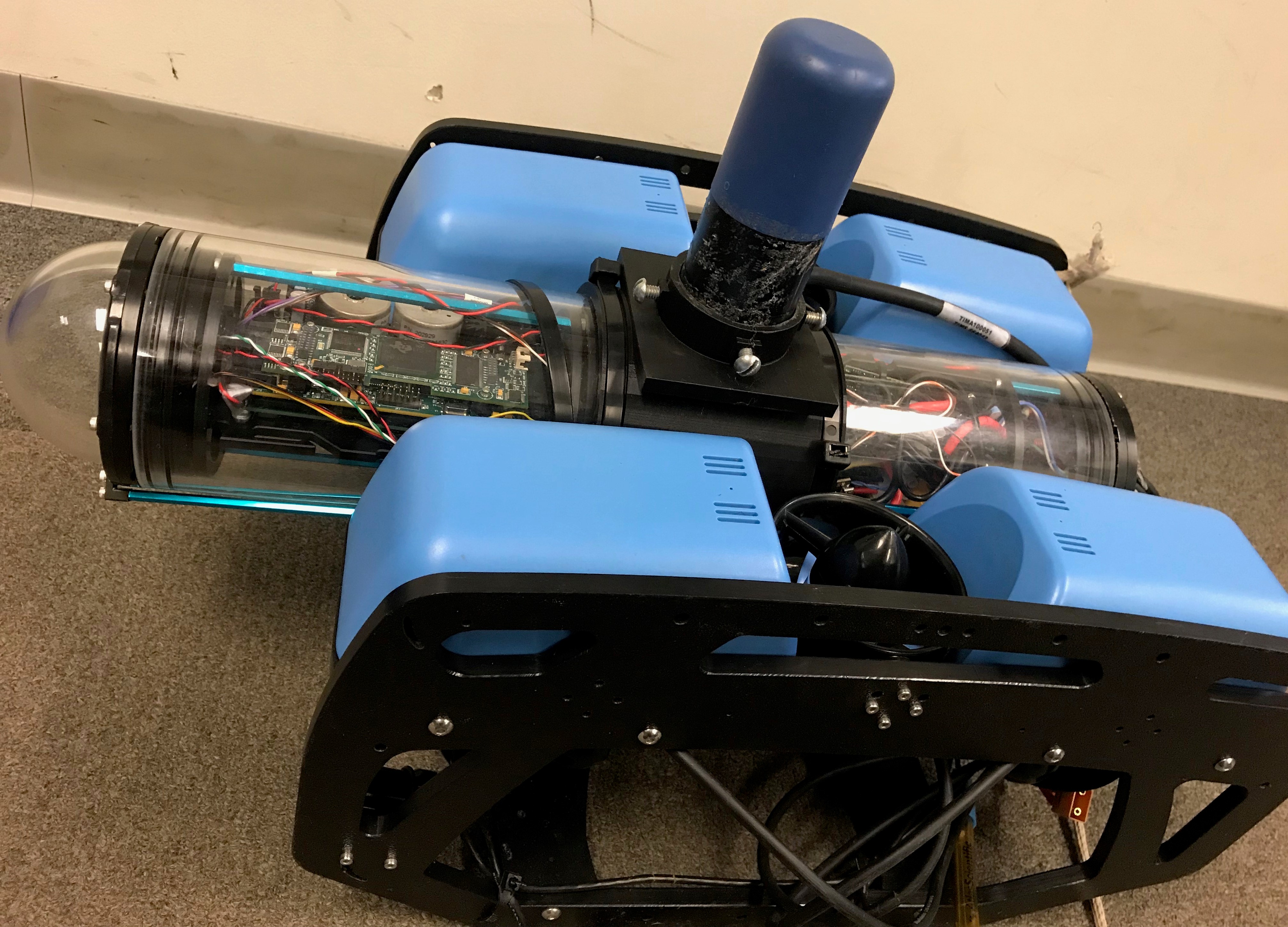}\\
\hspace{-0.5cm} (a)   \hspace{2cm} (b)  \hspace{2.5cm} (c)
\end{tabular}
\caption{(a)~Stack of boards connected to WHOI micro-modem rack; (b)~Stack of extrapolated boards in plastic stand; (c)~Installed micro-modem on vehicle.}\label{fig:boardinrack}
\end{figure}

The goal is to replace the default tethered connection of the ROV with an acoustic wireless connection to alter the missions far from shore as opposed to a fully autonomous vehicle so that the limitations of the tether (distance and hazards) do not limit the vehicle as with the current tethered connection. This will be accomplished using two WHOI Micro-Modems with  omnidirectional BTech BT25-UF $25~\rm{kHz}$ transducers.

\begin{figure}[!t]
\centering
\begin{tabular}{ccc}
\hspace{-3mm}
\includegraphics [width=0.118\textwidth]{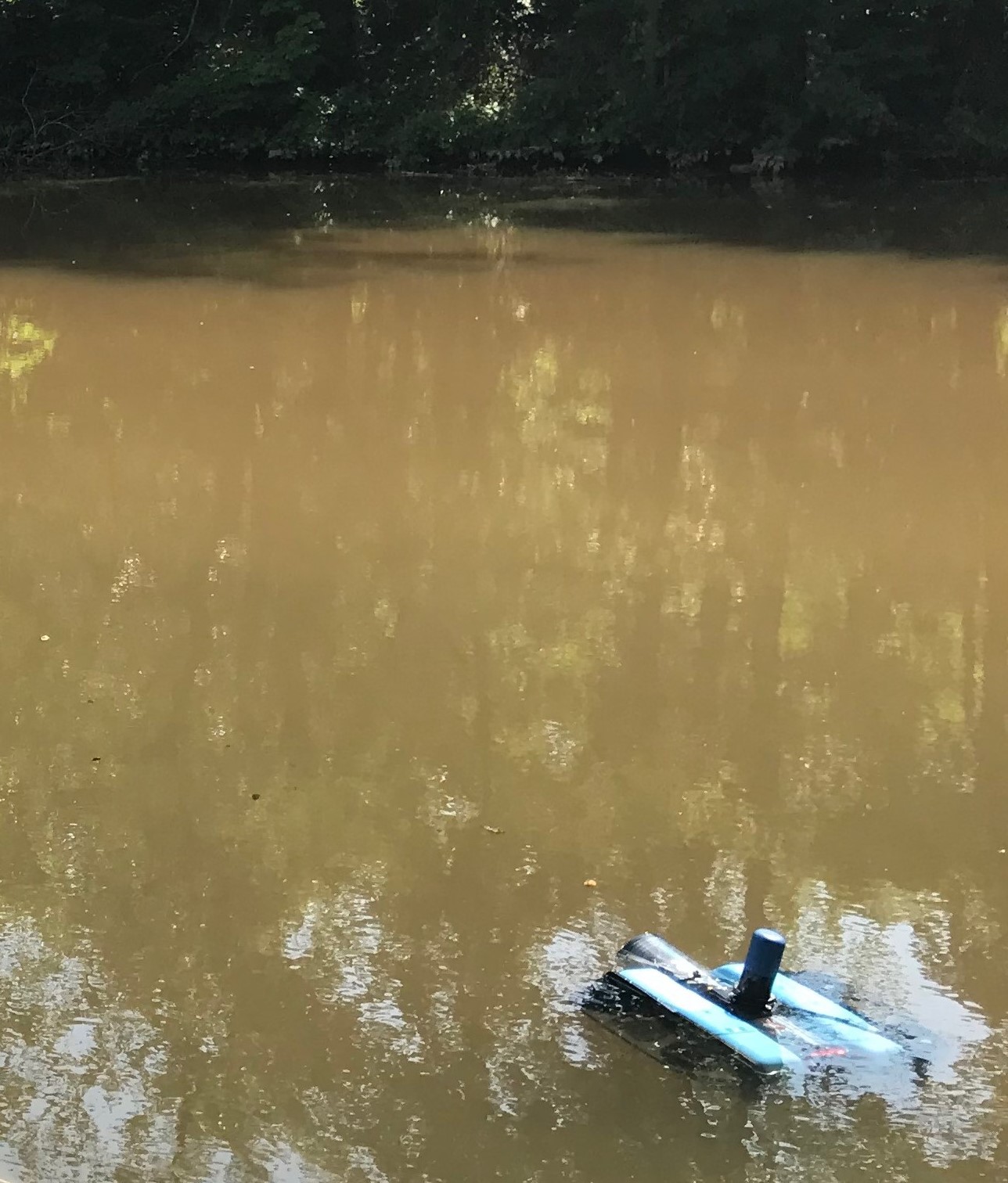}
\includegraphics [width=0.2\textwidth]{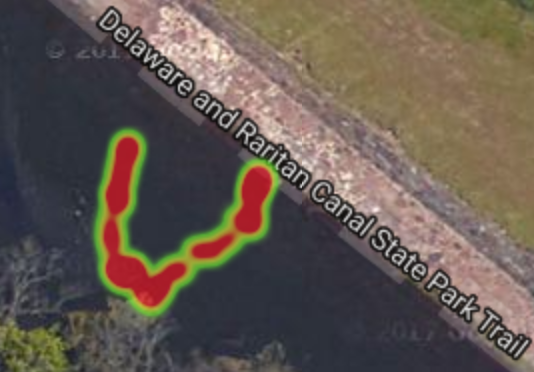}
\includegraphics [width=0.17\textwidth]{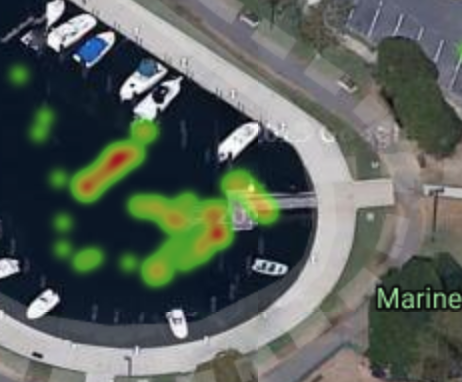}\\
\hspace{-0.7cm} (a)   \hspace{2.3cm} (b)  \hspace{2.5cm} (c) 
\end{tabular}
\caption{(a)~The vehicle in the Raritan river, NJ; Heatmap displayed from temperature data collected in (b)~Raritan river in Somerset, NJ; (c)~Marine park in Red Bank, NJ.}\label{fig:heatmap}
\end{figure}

We present a solution to fit and insert the micro-modems into the ROV electrical compartment and a way to attach the hydrophone onto the vehicle. The block diagram for shore to vehicle communication is shown in Fig.~\ref{fig:block}. The included parts are the co-processor board, micro-modem Digital Signal Processing~(DSP) board, power amplifier board, and hydrophone along with the custom-built cable connectors to connect to the hardware. While these seem like trivial problems to solve, with each change to the physical structure of the vehicle come consequences in terms of a completely different drag profile, weight distribution, and overall design of the motors.

To accommodate the stack of micro-modems, vehicle's enclosure has to be extended by $15~\rm{cm}$. The installation procedure is shown in Fig.~\ref{fig:boardinrack}(a)-(b). A clear acrylic cast tube of the proper dimensions replaces the original tube. The hydrophone is attached to the vehicle using a three-pin transducer cable with an Impulse LPMBH-3-FS watertight connector. In the enclosure, the cable is connected to a $24$ AWG wire, which is then twisted and fed into a Molex MicroFit $43645-0400$ connector in the micro-modem. The serial connection is a $6$ prong connector that connects to the micro-modem DSP board and on the other side connects to a small USB breakout board, which is then connected to a Raspberry Pi to communicate using NMEA serial commands. A $14.8~\rm{v}$ battery powers the ROV; the power is wired directly to terminals on the amplifier board. Fig.~\ref{fig:boardinrack}(c) shows the installed acoustic micro-modem on the vehicle, i.e., BlueROV2, and Fig.~\ref{fig:heatmap}(a) shows the vehicle moving in the Raritan river, NJ.

\begin{figure}[!t]
\centering
\begin{tabular}{cc}
\hspace{-5mm}
\includegraphics [width=0.24\textwidth]{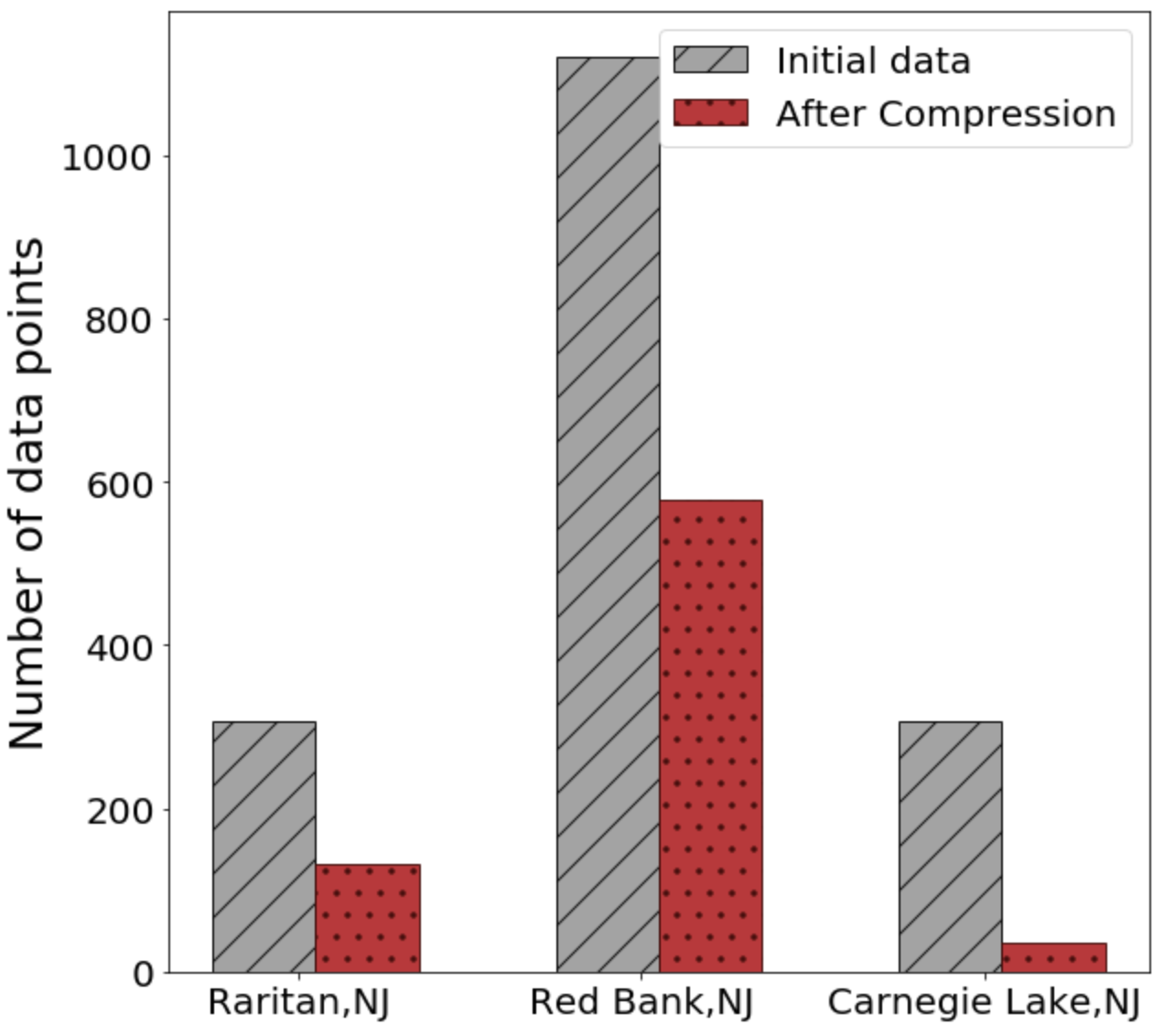}
\includegraphics[width=4.8cm,height=3.85cm]{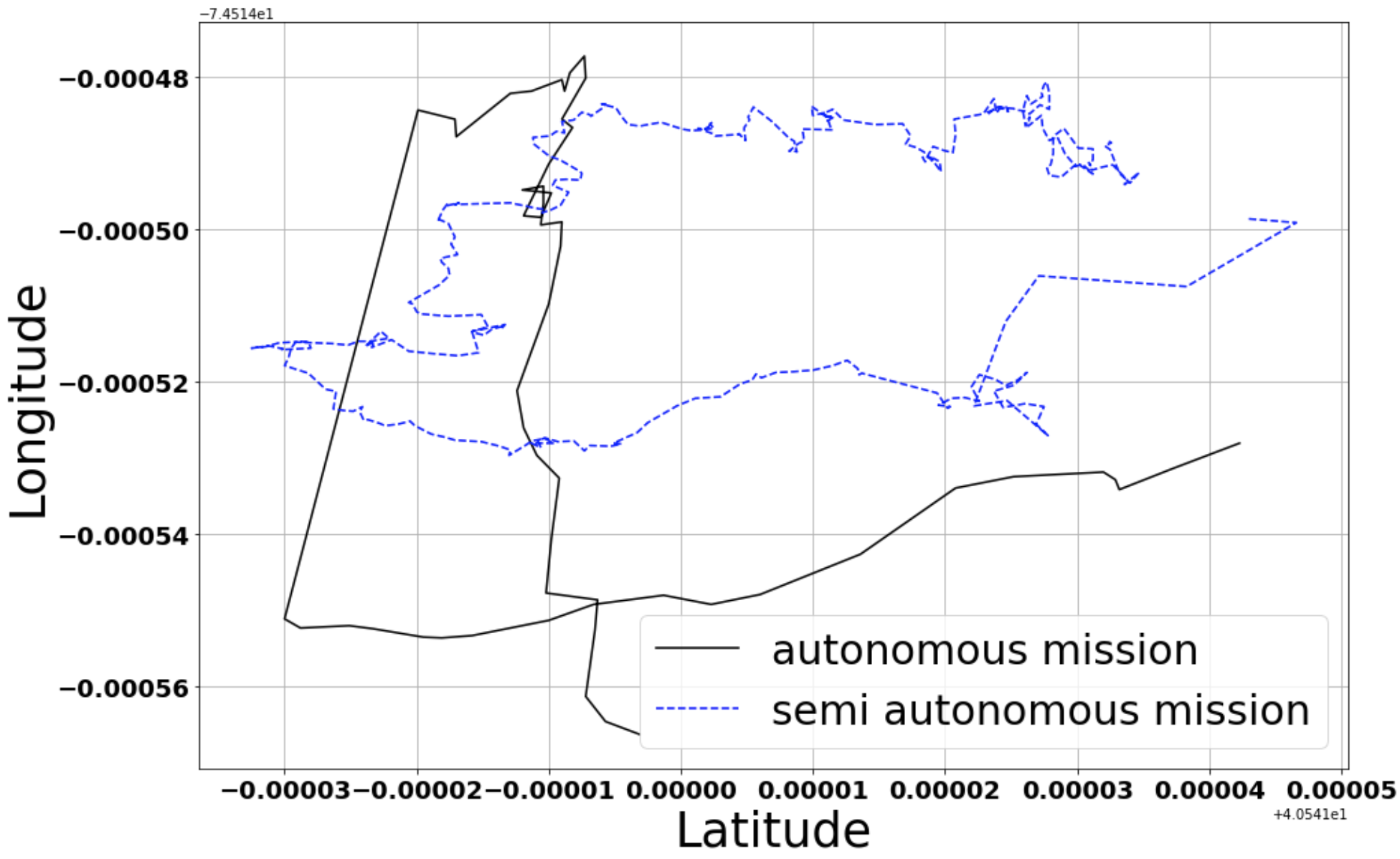}\\
\hspace{0.5cm} (a)   \hspace{4cm} (b) 
\end{tabular}
\caption{(a)~Compressed temperature data points; (b)~Square trajectory obtained when the vehicle is operated in autonomous and semi-autonomous modes in the Raritan river.}\label{fig:compressedtemp}
\vspace{-3mm}
\end{figure}

\begin{figure}[!t]
\centering
\hspace{-3mm}
\begin{tabular}{cc}
\includegraphics [width=0.24\textwidth]{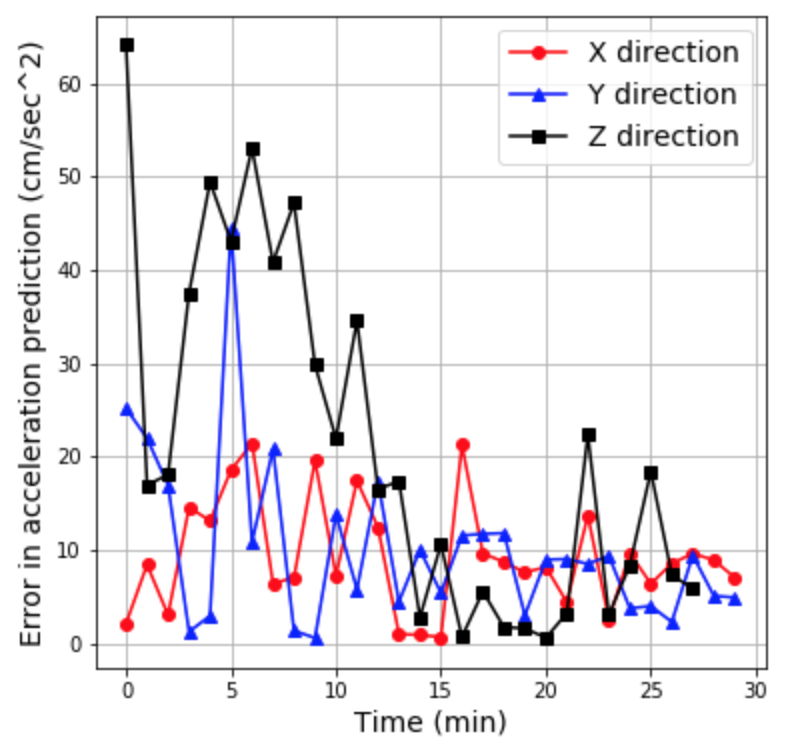}
\includegraphics [width=0.24\textwidth]{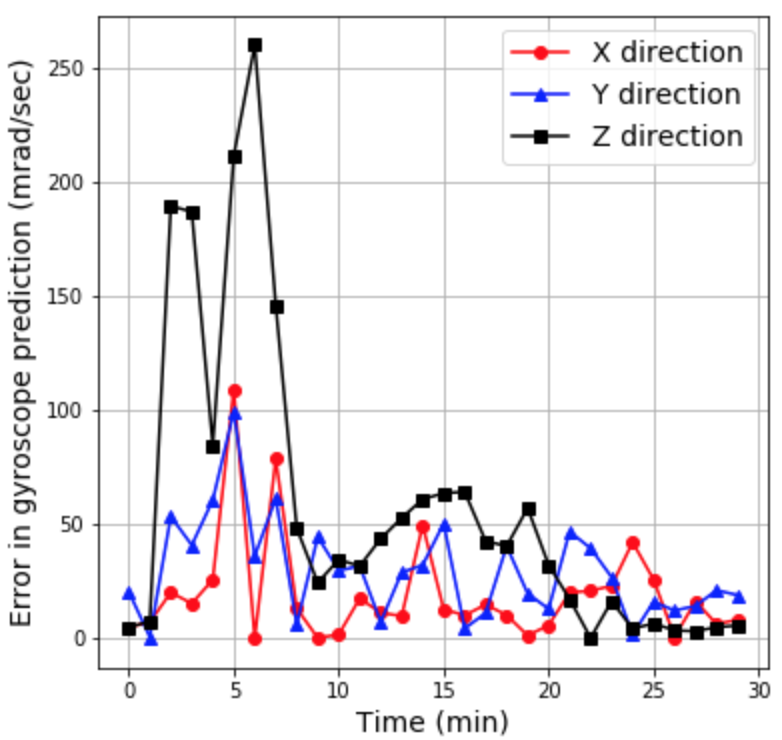}\\
\hspace{0.5cm} (a)   \hspace{4cm} (b) 
\end{tabular}
\caption{Error between the actual measurements of (a)~acceleration; (b)~gyroscope and the kalman Filter predictions.}\label{fig:error}
\vspace{-3mm}
\end{figure}

\textbf{Field Experiments:}
Experiments were conducted in the Raritan river, Princeton Carnegie Lake, and Marine Park in Red Bank, NJ during summer 2019. The BlueROV2 vehicle moved along a certain trajectory and samples of data have been collected. We extracted the temperature data and created heatmap to observe the change in temperature at various locations. This helps the controller study trends in the data and make decisions accordingly. Fig.~\ref{fig:heatmap}(a) and \ref{fig:heatmap}(b) show the temperature recorded by the BlueROV2, which is used for the purpose of semi-autonomous mission in the Raritan river and Marine park in Red Bank, NJ, respectively.

For the acoustic communication, we compressed the temperature data packets so that only relevant and unpredictable data is transmitted to the control center. Algorithm 1 have been applied in order to compress and retrieve certain data points that possess greater mean squares error in estimation. Those points are considered to be unpredictable. 
The regression analysis takes the cutoff value as the base point and chooses those points that are greater than the error threshold. The resultant compressed data points are represented in Fig.~\ref{fig:compressedtemp}(a) which 
shows that there are less compressed points in the Carnegie lake since most of the temperature data does not show huge differences. The temperature in redbank area as 
have the most unpredictable data points that are sent back to the control center to make decision in semi-autonomous mission. The decisions include going to the places that have higher or lower temperature. 

The IMU data has been compressed using model based approach described in Algo.~\ref{algo:compress}. Fig.~\ref{fig:error} compares the data measured by  of in different directions with the predictions obtained from the kalman filter. The mismatch points are considered unpredictable and transmitted through acoustic modems to the shore. 
The mismatch data that have been pointed out using kalman filter helps in trajectory missions as shown in Fig.~\ref{fig:compressedtemp}(b).

\section{Conclusion and Future Work}\label{sec:conc}
We modified an underwater ROV through the installation of a WHOI micro-modem to enable semi-autonomous missions for the vehicle through acoustic communications with a control center. In the proposed communications protocol, both vehicle and the center keep silent when the measured data matches the prediction. The transmission occurs only when unpredictable data is collected. We presented an efficient data-compression method to send the unpredictable data through acoustic communications. As future work, within the mode of control center-assisted autonomy in a multi-vehicle scenario, we will split up the functionalities such that the center is only responsible of mission restructuring and of some critical/safety functions, while the vehicles cooperatively accomplish team-determined goals.



\textbf{Acknowledment:}
This work was supported by the NSF CPS Award No.~1739315. The authors thank Tomasz Brzyzek, ECE grad student, as well as Agam Modasiya and Karun Kanda, MAE and CS UG students, respectively, for their help with the experiments.

\newpage

\bibliographystyle{ACM-Reference-Format}
\bibliography{sample-base,ref}

\appendix

\end{document}